\documentclass[aps,pre,preprint,showpacs,amssymb]{revtex4}
\usepackage{amsmath,graphics,epsfig}
\newcommand{\lb} {\makebox{{\Large [}}}
\newcommand{\rb} {\makebox{{\Large ]}}}
\newcommand{\obs }{\ensuremath{{\cal O}}}

\begin{document}
\title{Single-cluster dynamics for the random-cluster model}
\author{Youjin Deng~$^1$, Xiaofeng Qian~$^2$, 
Henk W.J. Bl\"ote~$^{2,3}$} 
\affiliation{
$^{1}$Hefei National Laboratory for Physical
Sciences at Microscale and Department of Modern Physics, 
University of Science and Technology of China, Hefei 230027, China \\
$^{2}$ Lorentz Institute, Leiden University,
  P.O. Box 9506, 2300 RA Leiden, The Netherlands \\
$^{3}$Faculty of Applied Sciences, Delft University of
Technology,\\  P.O. Box 5046, 2600 GA Delft, The Netherlands}
\date{\today} 

\begin{abstract} 
We formulate a single-cluster Monte Carlo algorithm for the simulation
of the random-cluster model. This algorithm is a generalization
of the Wolff single-cluster method for the $q$-state Potts model  to
non-integer values $q>1$.  Its results for static quantities are in a
satisfactory agreement with those of the existing
Swendsen-Wang-Chayes-Machta (SWCM) algorithm, which involves a full
cluster decomposition of random-cluster configurations. 
We explore the critical dynamics of this algorithm
for several two-dimensional Potts and random-cluster models. 
For integer $q$, the single-cluster algorithm can be reduced to the
Wolff algorithm, for which case we find that the autocorrelation
functions decay almost purely exponentially, with dynamic exponents
$z_{\rm exp} =0.07~(1),~0.521~(7)$, and $1.007~(9)$ for $q=2,~3$, 
and $4$ respectively. For non-integer $q$, the dynamical behavior
of the single-cluster algorithm appears to be very dissimilar
to that of the SWCM algorithm. For large critical
systems, the autocorrelation function displays a range of power-law
behavior as a function of time. The dynamic exponents are relatively
large. We provide an explanation for this peculiar dynamic behavior.
\end{abstract}
\maketitle 

\section { Introduction }
\label{sec1}
The Kasteleyn-Fortuin mapping \cite{KF} of the $q$-state Potts
model \cite{Potts} onto the random-cluster model provides a way
to define the Swendsen-Wang \cite{SW} and related cluster  Monte
Carlo algorithms \cite{UW,BC} for the Potts model. These algorithms
can apply nonlocal changes to the configuration. For systems with
long-range correlations, these nonlocal methods appear to be very
efficient in comparison with the standard Metropolis Monte
Carlo method \cite{NM} which applies only local updates.

The number $q$ of Potts states appears as a continuous variable in
the random-cluster model; the latter model can thus be seen as a
generalization of the Potts model to non-integer values of $q$.
There exist several ways to simulate non-integer-$q$ random-cluster
models. First, Sweeny applied local updates of the bond variables 
\cite{MS}. While the Sweeny algorithm, like cluster algorithms,
suppresses most of the critical-slowing down, a bond update requires 
a nonlocal task which increases the computer time requirements. 
Another algorithm
was formulated by Hu \cite{CKH}. It generates percolation configurations
and applies a statistical reweighting in order to obtain the correct
averages for the random-cluster model. A cluster algorithm of the
Swendsen-Wang type was formulated by Chayes and Machta \cite{CM} for
the $q\geq 1$ random-cluster model. While all these algorithms,
if using a random generator of a sufficient quality,
lead to results that are only subject to statistical errors, it
was found that the Swendsen-Wang-Chayes-Machta (SWCM) cluster 
algorithm (where applicable, i.e. $q>1$)
is much more efficient than the reweighting method  \cite{QDB}. It
was also found \cite{QDB} to be more efficient than the Sweeny method,
although the latter result depends strongly on
the sophistication of the bond update method.
Detailed studies of the dynamic critical behavior of 
the SWCM and the Sweeny algorithms can also be found 
in Ref.~\onlinecite{Deng}.

In this work, we present a single-cluster algorithm for the 
random-cluster model with real $q>1$. This algorithm thus has elements
in common with the Wolff \cite{UW} as well as with the
SWCM cluster algorithm~\cite{CM}. Since, for integer $q$, the Wolff
method is about as efficient as the Swendsen-Wang algorithm, one might
expect that the same holds for our single-cluster algorithm in 
comparison with the SWCM algorithm. A test of this expectation 
is also included in the present work.

Section \ref{sec2} provides an explanation of the theoretical aspects
of the new algorithm. We include a simple example of such an algorithm,
and prove that the condition of detailed balance is satisfied.
Furthermore, we describe other variants of the algorithm that are
applicable to models with $q>2$, and describe how the algorithm can
reduce to the Wolff algorithm for integer $q$. In Sec.~\ref{sec3} we
test the validity of the algorithm, and determine its dynamic exponent
for two-dimensional random-cluster models on the square lattice, using
several values of $q$. The generic dynamical behavior appears to be very
different from that of the Wolff algorithm. We conclude with a discussion
of our findings in Sec.~\ref{sec4}, which also includes an explanation of
the mechanism responsible for the unusual dynamical behavior.

\section {Algorithm }
\label{sec2}
\subsection {The Kasteleyn-Fortuin mapping }
We recall the mapping of the Potts model and the random-cluster model
on a model with bond as well as site variables, as described in
Refs.~\onlinecite{CM} and \onlinecite{QDB}.
For a review of the Potts model, see, e.g. Ref.~\onlinecite{FYW}.
The Potts partition sum is
\begin{equation}
Z_{\sigma} \equiv \lb \prod_{i=1}^{N} \sum_{\sigma_i=1}^q\rb
 \prod_{\langle ij \rangle} \exp(K \delta_{\sigma_i\sigma_j})\, ,
\label{Zpo}
\end{equation}
where the summations are on all site variables $\sigma_i$, where $i$
labels the lattice sites.  The second product sign indicated by
$\langle ij \rangle$  is on all nearest-neighbor pairs $(i,j)$.
The coupling $K$ is reduced, i.e., it includes a factor $1/k_BT$.
We consider the case of ferromagnetic couplings $K\geq 0$.
The Kasteleyn-Fortuin mapping of Eq.~(\ref{Zpo}) on the random-cluster
model \cite{KF} introduces bond variables $b_{ij}=0$ or 1 between
all neighbor pairs $(i,j)$, after which the site variables
$\sigma_i=1,2,\cdots,q$ can be summed out so that only the bond
variables remain as the degrees of freedom of the random-cluster
model. Bonds $b_{ij}=1$ (0) are considered to be present (absent).

The random-cluster partition sum  thus assumes the form
\begin{equation}
Z_{\sigma}=Z_{b} \equiv
\lb \prod_{\langle ij \rangle} \sum_{b_{ij}=0}^1\rb q^{n_c} u^{n_b}
=\sum_{\{b\}}\prod_{k=1}^{n_c} q u^{n_{b}^{(k)}}\, ,
\label{Zrc}
\end{equation}
where $u\equiv {\rm e}^K-1$ is the temperature-like parameter,
$n_b\equiv \sum b_{ij}$ denotes the number of present bonds. The
number of clusters is denoted as $n_c$. The sum on ${\{b\}}$
is shorthand for the sum on all configurations of bond variables, and
$n_{b}^{(k)}$ is the number of nonzero bonds in the $k$-th cluster.

For $q>1$ one can divide the cluster weight $q$ in two positive
contributions 1 and $q-1$. The first contribution can be associated
with one of the original Potts states. To this purpose we introduce
`color' variables $\tilde{t}_k=0$ or 1 for each cluster $k=1,2,\cdots,n_c$:
\begin{equation}
Z_{b} = \sum_{\{b\}} \prod_{k=1}^{n_c} \sum_{\tilde{t}_k=0}^{1}
u^{n_b^{(k)}\tilde{t}_k} \lb (q-1)u^{n_b^{(k)}}{\rb}^{1-\tilde{t}_k} \, .
\label{Ztt}
\end{equation}
Clusters of color 0 and 1 have weight $q-1$ and 1 respectively. The sum
on the cluster colors is replaced by a sum over $N$ site-color variables
$t_i=0$ or 1, together with a factor $\delta_{t_it_j}^{b_{ij}}$
(with the convention $0^0$=1) for each bond variable, so that each
cluster contains only sites of one color:
\begin{equation}
Z_{b}=Z_{tb} \equiv \sum_{\{t\}} \sum_{\{b\}} \prod_{\langle ij \rangle}
(u \delta_{t_it_j})^{b_{ij}} \prod_{k=1}^{n_c} (q-1)^{1-t_{s(k)}}\, ,
\label{Ztb}
\end{equation}
where the color of the $k$-th cluster is denoted $t_{s(k)}$ where
$s(k)$ is a site in that cluster. In a site 
configuration $\{t\}$ we identify 3 types of bonds $(ij)$:
\begin{eqnarray}
{\rm type}\; 0 &:& t_i=t_j=0~;   \nonumber \\
{\rm type}\; 1 &:& t_i=t_j=1~;   \nonumber \\
{\rm type}\; 2 &:& t_i+t_j=1~.   \nonumber 
\end{eqnarray}
Summations and products involving only one of these types of bond
are specified by appending corresponding superscripts to the pertinent
summation and product signs:
\begin{displaymath}
Z_{tb} = \sum_{\{t\}}
\lb {\sum_{\{b\}}}^{(0)} {\prod_{\langle ij \rangle}}^{(0)} u^{b_{ij}}\rb
                                      \lb \prod_{k=1}^{n_c^{(0)}}(q-1)\rb
\lb {\sum_{\{b\}}}^{(1)} {\prod_{\langle ij \rangle}}^{(1)} u^{b_{ij}}\rb
\end{displaymath}
\begin{equation}
\lb {\sum_{\{b\}}}^{(2)} {\prod_{\langle ij \rangle}}^{(2)} (1-b_{ij})\rb\, ,
\label{Ztb3}
\end{equation}
where the clusters of color 0 are labeled $1,2,\cdots,n_c^{(0)}$.
The type 1 and 2 sums can now be executed. After rewriting the type-0
sum one obtains the partition sum expressed in site variables and 
type 0 bond variables:
\begin{equation}
Z_{tb} = Z_{tb1} \equiv  \sum_{\{t\}} 
{\sum_{\{b\}}}^{(0)} 
\lb  {\prod_{k=1}^{n_c^{(0)}}} (q-1)u^{n_b^{(k)}}\rb
\lb  {\prod_{\langle ij \rangle}}^{(1)}  (1+u)\rb \, .
\label{Ztb1}
\end{equation}
Eq.~(\ref{Ztb1}) specifies the probability distribution of a system
of site variables $t_i=0,1$ and bond variables $b_{ij}$ between
nearest-neighbor sites of type 0. 
Each term in the second sum in Eq.~(\ref{Ztb1}) specifies a cluster
decomposition ${\mathcal D}(\{b\})$ of the sublattice formed by
sites $k$ with $t_k=0$.
Different sets of bond variables $\{b\}$ may still correspond
with the same cluster decomposition. Thus, if we replace the sum on
$\{b\}$ by a sum on all cluster decompositions of the color-0 regions,
we have to insert a
summation on all  $\{b\}$ that are consistent with ${\mathcal D}$:
\begin{equation}
Z_{tb} = Z_{t{\mathcal D}} \equiv  \sum_{\{t\}}
\lb  {\prod_{\langle ij \rangle}}^{(1)}  (1+u) \rb
{\sum_{\{{\mathcal D}\}}}^{(0)}
{\sum_{\{b\}|{\mathcal D}}}
\lb  {\prod_{k=1}^{n_c^{(0)}}} (q-1)u^{n_b^{(k)}}\rb \, .
\label{ZtD}
\end{equation}

\subsection {The simplest form of the algorithm }
\label{simpal}
Eq.~(\ref{ZtD}) can serve as the basis on which a single-cluster
Monte Carlo algorithm can be constructed.
This algorithm is applied as follows to a mixed configuration 
specified by the site variables $t_i$ and a cluster decomposition 
${\mathcal D}$ of the color-0 sites. An initial configuration can,
for instance, be obtained from a random-cluster configuration and
assigning color 1 to each cluster with probability $1/q$. 
Then, a single-cluster step is executed as follows:
\begin{description}
\item[1]
Choose a random site $i$. The action taken by the algorithm depends 
on the color variable $t_i$. If 
\item[2a]
$t_i=1$, do with probability $p_1=(q-1)/q$ the following:
form a random cluster around
site $i$ with bond probability $p=u/(u+1)$ between sites of color 1.
The sites $j$ in the newly formed cluster are assigned color 0
(i.e. $t_j=0$) and the number $n_c^{(0)}$ of clusters of color 0
is thus increased by 1.
\item[2b]
$t_i=0$, do with probability $p_2=1/q$ the following:  
assign color 1 to all sites of the cluster containing site $i$, and
thus decrease the number of clusters of color 0 by 1.
\end{description}

\subsection {Proof of detailed balance }
The proof of detailed balance can be formulated as follows.
Consider two mixed configurations $S_1$ and $S_2$, which differ only
in a region ${\mathcal C}$ whose sites $j$ have color $t_j=1$ in $S_1$,
and whose sites belong to a single cluster in $S_2$, and thus have 
color $t_j=0$. According to the rules given in the preceding subsection,
the transition probability to move from $S_1$ to $S_2$ is
\begin{equation}
T(2,1) = \frac{(q-1)N_c}{qN} \sum_{\{b\}|{\mathcal C}} 
\left( \frac{u}{u+1}\right)^{n_b}
\left( \frac{1}{u+1}\right)^{n_p+n_{nn}-n_b}\, ,
\label{T21}
\end{equation}
where $N_c$ is the number of sites in region ${\mathcal C}$; $N$ is the
total number of sites in the system; $\{b\}$ stands for the $n_{nn}$
bond variables on the edges between nearest-neighbor sites in ${\mathcal C}$; 
the combination on $\{b\}|{\mathcal C}$ indicates the sum on all
configurations $\{b\}$ that connect all sites in ${\mathcal C}$ into a
single cluster; $n_b$ denotes the number of nonzero bond variables in
$\{b\}$; $n_{p}$ is the number of bond variables connecting sites
sites inside ${\mathcal C}$ with those outside ${\mathcal C}$ of
color 1 (i.e., the number of bonds along the boundary of ${\mathcal C}$
that is broken when the color of ${\mathcal C}$ is changed).
The prefactor $(q-1)N_c/qN$ describes the probability that the cluster
formation starts within ${\mathcal C}$. Each of the $n_b$ `present' bonds 
contributes a factor $u/(u+1)$, and each of the $n_{nn}-n_b$ `absent'
bonds a factor $1/(u+1)$. Also each `broken' bond along the perimeter
of ${\mathcal C}$ contributes a factor $1/(u+1)$.

The rules given in the preceding subsection also specify the probability
of the inverse move, namely from $S_2$ to $S_1$, as 
\begin{equation}
T(1,2) = \frac{N_c}{qN} \, .
\label{T12}
\end{equation}
The condition of detailed balance requires that the transition
probabilities $T(2,1)$ and $T(1,2)$ are related to the equilibrium
probabilities $P(1)$ and $P(2)$ of configurations 1 and 2 respectively:
\begin{equation}
T(2,1) / T(1,2) = P(2) / P(1) \, .
\label{db}
\end{equation}
Since the probabilities $P(1)$ and $P(2)$ are proportional to the
configuration weights specified by Eq.~(\ref{ZtD}), we may write
\begin{equation}
P(2) / P(1) = W(2) / W(1)  \, ,
\end{equation}
where the weights associated with region ${\mathcal C}$ in
Eq.~(\ref{ZtD}) are
\begin{equation}
W(1) = (1+u)^{n_p+n_{nn}}
\label{W1}
\end{equation}
and 
\begin{equation}
W(2) = (q-1) \sum_{\{b\}|{\mathcal C}} u^{n_b} \, .
\label{W2}
\end{equation}
From Eqs.~(\ref{T21}) and (\ref{T12}), and from  Eqs.~(\ref{W1}) and
(\ref{W2}), we conclude that
\begin{equation}
T(2,1) / T(1,2)= 
(q-1) (1+u)^{-n_p-n_{nn}} \sum_{\{b\}|{\mathcal C}} u^{n_b}=
 W(2) / W(1) \, ,
\end{equation}
which shows that the condition of detailed balance, Eq.~(\ref{db}),
indeed is satisfied.

\subsection {Other versions }
\label{over}
The probabilities $p_1$ and $p_2$ in Sec.~\ref{simpal} can be chosen
differently, depending on the value of $q$. For $1<q<2$ we may take
$p_1=q-1$ and $p_2=1$. For $q>2$, this is not possible but other
possibilities arise. One can generalize the algorithm by allowing
more than two values of the color variables $t_i$.
The most obvious way is to allow $n\equiv[q]$ (the integer part of $q$)
values with weight one, and one special value with weight $q-n$.
Sites of the latter color are divided in clusters (just as before);
sites of the $n$ remaining colors are not. A cluster step starting
from a randomly chosen site can then be specified as follows: if
that site belongs to a cluster (thus, of the special color 0), then the
cluster is erased and its sites are given a random color $1,2,\cdots,n$
with probability $1/n$ each. If the cluster step starts from a randomly
chosen site of color $1,2,\cdots,n$, then a single cluster is formed.
Its sites receive one of the $n-1$ other weight-1 colors with probability
$(2n-q)/[n(n-1)]$ each, and the cluster receives the special color with
probability $(q-n)/n$.
This choice of probabilities satisfies detailed balance and maximizes
the probability of a cluster flip. For integer $q$ it reduces to the
Wolff algorithm.

\subsection { Test of the algorithm}
We tested the single-cluster algorithm for the cases of the $q=2$, $3$,
and $4$ Potts model on the square lattice, by comparing its numerical
results to those of the Wolff algorithm. We set $n=q-1$ 
(see Sec.~\ref{over}) and the weight of the color-0 clusters is thus
$q-n=1$.  Simulations were performed on
$L \times L$ lattices with periodic boundary conditions.
After each single cluster step, we sampled various quantities,
including the densities $\rho_i$ of Potts variables in states 
$i=1,2,\cdots,q$, and the single cluster size $S$.
The single cluster size is counted as the total number of lattice
sites in the cluster as constructed by the algorithm.  If the number
$q$ of Potts states is an integer, the squared Potts magnetization 
density $m^2$ can be expressed in the densities $\rho_i$  as
\begin{equation}
m^2 = \frac{1}{q-1} 
\sum_{i} \sum_{j < i} (\rho_i-\rho_j)^2  
=\frac{q}{q-1}\sum_{i} (\rho_i-1/q)^2 \, .
\end{equation}
The sum on the right-hand side of this equation contains $q$ terms whose
expectation value is equal, due to the Potts symmetry. Thus, for the
expectation value $\langle m^2 \rangle$ of $m^2$ we may write
\begin{equation}
\langle m^2 \rangle=\frac{q^2}{n(q-1)}
\sum_{i=1}^{n} \langle (\rho_i-1/q)^2 \rangle \, ,
\label{msamp}
\end{equation}
with $1 \leq n \leq q$.
Thus it is sufficient to sample $(\rho_i-1/q)^2$ in order
to obtain $\langle m^2 \rangle$. While $q$ is taken to be an integer
in this subsection, Eq.~(\ref{msamp}) still applies for general $q>1$.  
If $q$ is not an integer, $n$ will usually be chosen as $n=[q]$ where
$[q]$ denotes the integer part of $q$. Although, in the case $n<q$, 
Eq.~(\ref{msamp}) still leads to the same expectation values as those
obtained by averaging on the basis of a full cluster decomposition,
the autocorrelations of $m^2$ may depend on the sampling method and
thus be different in both cases.

As should be expected, for Potts models with integer values of $q$,
the Wolff and the present algorithm did indeed yield mutually consistent
results. This is demonstrated by the data for $\langle m^2 \rangle$ and
$\langle S \rangle$ in Table~\ref{tab_1} obtained by the two algorithms for
the case $q=2$, $n=1$. Furthermore, since the probability to hit a cluster 
is equal to its relative size, it follows that the two expectation values
$\langle m^2 \rangle$ and $\langle S \rangle$ are equal. Our simulation
results were also in a good agreement with this relation, as illustrated
by the data shown in Table~\ref{tab_1} for the critical Ising model. 

Since both simulations involve the same number of samples,  
the statistical uncertainties, shown between brackets
in Table~\ref{tab_1}, reflect the relative efficiency of the Wolff 
and the single-cluster algorithm. For size $L=8$, the Wolff method
is about $10$ times as efficient as the present algorithm, while
this difference increases to a factor of about 100 for $L=32$.
It thus appears that the two algorithms have different dynamic exponents.

\section { Dynamic exponents}
\label{sec3}

\subsection{Autocorrelation functions and autocorrelation times}
\label{autocorrelation times}
Consider an observable $\obs$, whose evolution in time $t'$ is described
by the time-series $\obs(t')$, where each unit of $t'$ corresponds to
one step of the single-cluster algorithm. The autocovariance function
of $\obs$ is defined to be
\begin{equation}
C_{\obs}(t') \equiv \langle \obs(0)\obs(t')\rangle-\langle\obs\rangle^2 \, ,
\end{equation}
and its autocorrelation function is
\begin{equation}
A_{\obs }(t') \equiv \frac{C_{\obs}(t')}{C_{\obs}(0)} \, .
\end{equation}
We then normalize time $t'$ as $t=t' {S}/L^2$ so that the time
unit of $t$ is the average number of cluster steps in which each lattice
site is visited once.
From $A_{\obs}(t)$ we then define the integrated
autocorrelation time as
\begin{equation}
\tau_{{\text{int}},\obs} \equiv \frac{1}{2}+\sum_{t=1}^{\infty}\,
A_{\obs }(t) \, ,
\label{tauint_definition}
\end{equation}
and the exponential autocorrelation time as
\begin{equation}
\tau_{{\text{exp}},\obs} \equiv \lim_{t\to\infty}
\frac{-t}{\log\,A_{\obs}(t)} \, .
\label{tauexp definition}
\end{equation}
Finally, the exponential autocorrelation time of the system is defined
as
\begin{equation}
\tau_{\text{exp}}=\sup_{\obs}\tau_{{\text{exp}},\obs} \, ,
\end{equation}
where the supremum is taken over all observables $\obs$. This
autocorrelation time measures the decay rate of the slowest mode of
the system. All observables that are not orthogonal to this slowest
mode satisfy $\tau_{{\text{exp}},\obs}=\tau_{\text{exp}}$.

\subsection {Integer $q$}
\label{intq}
For integer $q=2,3,4$, we may set $n=q$, in which case the color-0
clusters have zero weight and are thus absent, so that the
single-cluster algorithm reduces to the well-known Wolff
algorithm~\cite{UW}. Such Wolff simulations were performed at
criticality. The system sizes were  chosen as powers of 2 in the range
$4 \leq L \leq 4096$ for $q=2$, $4 \leq L \leq 2048$ for $q=3$,
and $4\leq L \leq 1024$ for $q=4$. Samples were taken at intervals of 
one single-cluster step. The number of samples taken for each system
size is shown in Table~\ref{tab_sample}.

\begin{figure}
\begin{center}
\leavevmode
\epsfxsize 10.6cm
\epsfbox{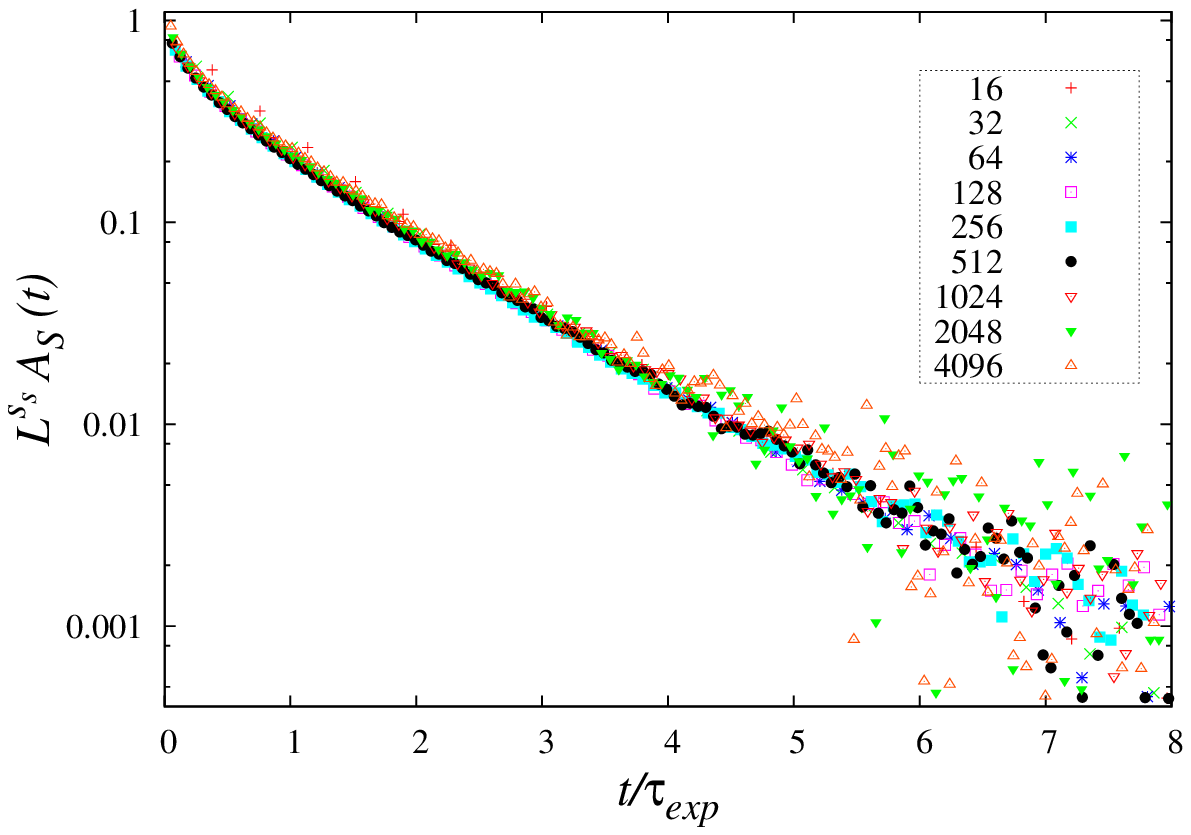}
\end{center}
\caption{(Color online) Data collapse of the autocorrelation function of
the single-cluster size, shown as $L^{s_s} A_{S}$ on a logarithmic scale,
versus  $t/\tau_{\rm exp}$, with $s_s=0.37$. These results apply to $q=2$
Wolff dynamics.} 
\label{fig_gwf2_00c}
\end{figure}

After a fast initial decay, the autocorrelation functions for $S$ and
$m^2$ decay approximately exponentially, but with an amplitude
proportional to a power of the linear size $L$. Except for the initial
decay, the behavior can be described as
\begin{equation}
A_{\obs} (t) \propto L^{-s_{\obs}} e^{-t/\tau_{\rm exp}(L)}  \, ,
\label{beh_rho_intq}
\end{equation}
with $\obs=S$ or $m^2$, which implies that
$z_{{\rm int},\obs}=z_{\rm exp} - s_{\obs} $.
Accordingly, a data collapse is obtained by plotting the quantity
$L^{s_s} A_{S}$ versus $t/\tau_{\rm exp}$. This is shown in
Fig.~\ref{fig_gwf2_00c}, with the exponent of $L$ fixed as $s_s=0.37$. 

Correlations between subsequent Wolff steps are thought to arise from
overlap between the two pertinent clusters. The average Wolff 
cluster size, relative with respect to the size $L^d$ of the system at
criticality, scales with $L$ as $S\propto L^{2y_h-2d}$,  where $y_h$
is the magnetic exponent and $d=2$ is the spatial dimensionality. The probability
that two subsequent clusters overlap may thus be crudely estimated as
$L^{4d-4y_h}$. The histogram of the cluster-size distribution is
however very wide with a large-size cutoff that scales as $L^{d-y_h}$.
Since large clusters contribute more to the autocorrelation function
than small ones, one may expect that the correlations at short times
scale with $L$ instead as $L^{-s_s}$ with $s_s<4d-4y_h$.
The results for $q=2$, 3, and 4 are shown in Table~\ref{tab_realq}.
It seems that for $q=4$ the Wolff algorithm is slightly less efficient 
than the Swendsen-Wang method.

\begin{figure}
\begin{center}
\leavevmode
\epsfxsize 10.6cm
\epsfbox{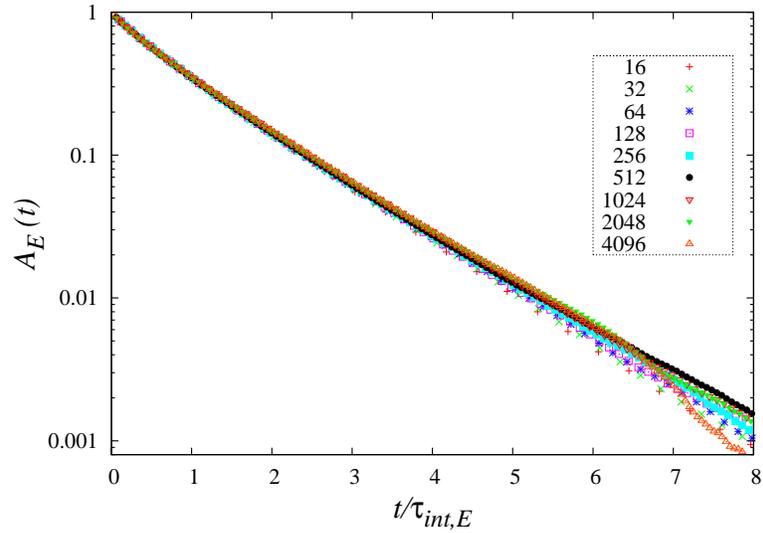}
\end{center}
\caption{(Color online) Data collapse of the autocorrelation function
$A_{E}$ vs. $t/\tau_{{\rm int}, {E}}$ for $q=2$ Wolff dynamics.
The system sizes $L$ are shown in the figure.
The statistical uncertainties become appreciable at large times.}
\label{fig_geb2_00b}
\end{figure}

\begin{figure}
\begin{center}
\leavevmode
\epsfxsize 10.6cm
\epsfbox{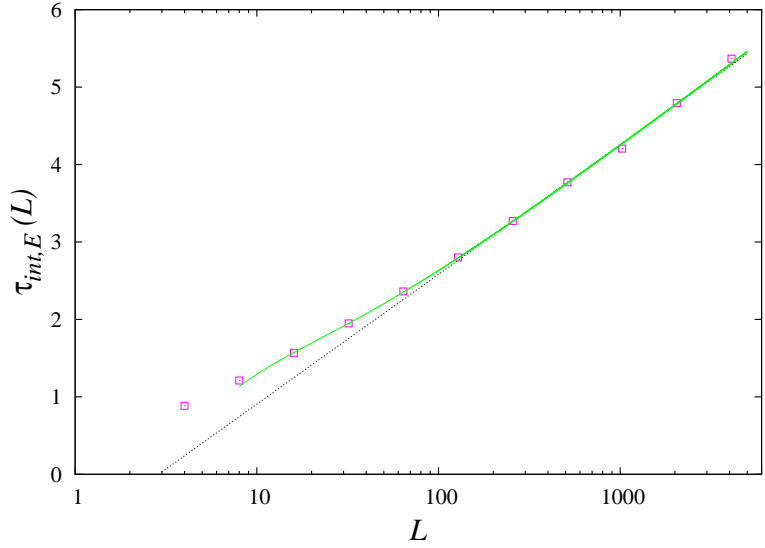}
\end{center}
\caption{(Color online) Semi-logarithmic plot of the integrated
autocorrelation function $\tau_{{\rm int},E} (L)$ versus $L$ for
$q=2$ Wolff dynamics. The error bars are of the same size as the 
data points. 
The solid (green) line is obtained from the logarithmic fit. The
difference with the power-law fit would not be visible on this scale.
The straight dashed line represents pure logarithmic behavior
$\tau \propto \ln L$, and serves only for the purpose of illustration.} 
\label{fig_geb2_00d}
\end{figure}

During the simulations, also the energy density $E$, which is defined as
the nearest-neighbor correlation function, was sampled. The corresponding
autocorrelation function $A_E(t)$ is shown in Fig.~\ref{fig_geb2_00b}
for $q=2$. This figure indicates that $A_{E}$ decays approximately
exponentially as a function of $t$, with an amplitude that has little
or no dependence on the system size. It thus follows that 
$z_{{\rm exp},E} \approx z_{{\rm int},E}$.
The autocorrelation times $\tau_{\rm int}$ and $\tau_{\rm exp}$ were
obtained by integration and least-squares fits respectively.
The autocorrelation times for $L \geq 16$ were fitted by
\begin{equation}
\tau_{{\rm int},E} (L) =a+b L^{z_{{\rm int},E}} \; ,
\label{fit_tau}
\end{equation}
and similarly for $\tau_{\rm exp}$. The fit yields
$z_{\rm exp} \approx z_{{\rm int},{E}} =0.07~(1)$. 
This nonzero dynamic exponent is in agreement with the upward curvature
of the data for $\tau_{\rm exp}$ versus $L$ on a logarithmic scale,
shown in Fig.~\ref{fig_geb2_00d}.
However, we cannot exclude the possibility that the dynamic exponent
is zero, because the data for $L \geq 16$ can also be described by
$\tau_{{\rm int},E} (L) = \tau_0 + \ln L (a_0 + a_1/L+a_2/L^2)$,
which has only one more parameter than Eq.~(\ref{fit_tau}), with
$\tau_0=-1.02 ~(7)$, $a_0=0.76 ~(2)$, $a_1 =3.4 ~(6)$, and $a_2=-10 ~(5)$;
this is illustrated in Fig.~\ref{fig_geb2_00d}.
Such behavior would mean that the Li-Sokal bound~\cite{LiS} is sharp for
the Wolff dynamics of the two-dimensional Ising model.

\subsection {Non-integer $q$}

We performed simulations of critical random-cluster systems with sizes
$ 4 \leq L \leq 256$ for $q=1.25$, 1.50, 1.75, 2.25,2.50, and $2.75$,
with $n=[q]$. Samples were taken after each single-cluster step, with a
total number of samples of $6 \times 10^7$ for each $L,q$. The squared
magnetization was obtained using Eq.~(\ref{msamp}).

\begin{figure}
\begin{center}
\leavevmode
\epsfxsize 10.6cm
\epsfbox{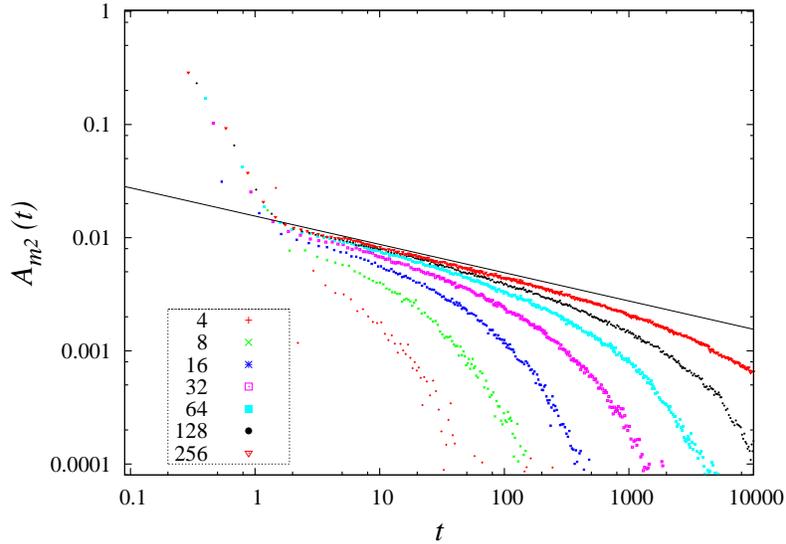}
\end{center}
\caption{(Color online) Autocorrelation function $A_{m^2}$ for 
$q=1.25$ versus time $t$,
using logarithmic scales. These data apply to the single-cluster 
simulation of the $q=1.25$ random-cluster model. The straight
line is only for the purpose of illustration, and has slope $-0.25$.}
\label{fig_gmt1_25a}
\end{figure}

The autocorrelation functions $A_{m^2} (t)$ and  $A_{S} (t)$ were
found to display a fast decay at short times $t \sim {\rm O} (1)$,
then decay algebraically, and ultimately exponentially with $t$.
Such a range of algebraic behavior, which extends to $t \gg 1$ for at
large $L$, is absent for Wolff dynamics. In the case
of $A_{m^2} (t)$, the fast initial decay at small $t$ appears to be 
hardly size-dependent, as can be seen in Fig.~\ref{fig_gmt1_25a}.
In contrast, for $A_{S} (t)$, the amplitude of the algebraic decay is
found to be size dependent. 

These dynamic phenomena are very different from those for integer $q$,
where the autocorrelation functions for both quantities decay
almost as a pure exponential law.  It seems that the
behavior of $A_{\obs} (t)$ can be described by
\begin{equation}
A_{\obs} (t,L) = L^{-s_{\obs}} t^{-r_{\obs}} f (t/\tau_{\rm exp} (L) )
\hspace{1cm} \mbox{for } t \gg 1 \;, 
\label{beh_rho_realq}
\end{equation}
where $f$ is a universal function. For large $t$, it behaves as 
\begin{equation}
f(t/\tau_{\rm exp} (L) ) \propto e^{-t/\tau_{\rm exp} (L)} \hspace{1cm}
\mbox{with } \tau_{\rm exp} (L) \propto L^{z_{\rm exp}} \; .
\end{equation}

We analyzed $A_{\obs}(t,L)$ by attempting to collapse the data 
onto a single curve according to Eq.~(\ref{beh_rho_realq}).
The data collapses for $\obs=m^2$ and $S$ work only approximately. 
This might be due to finite-size corrections. The results are shown in
Table~\ref{tab_realq}.

\begin{figure}
\begin{center}
\leavevmode
\epsfxsize 10.6cm
\epsfbox{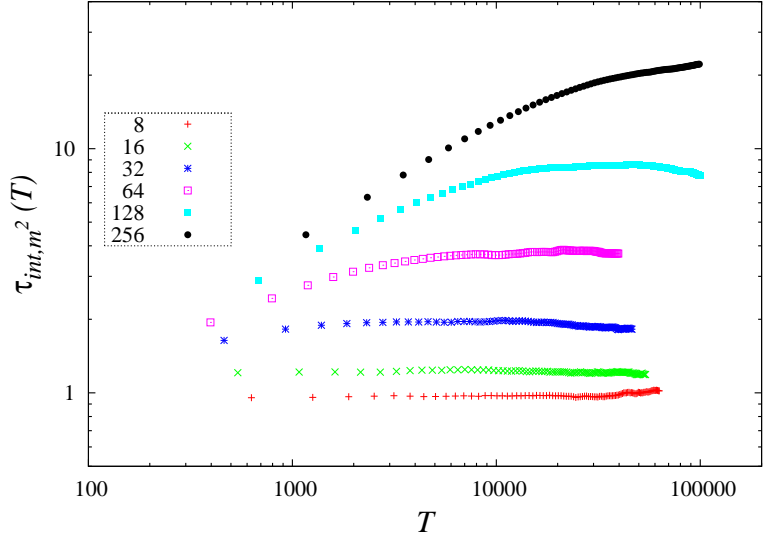}
\end{center}
\caption{(Color online) Indefinite integral $\tau_{{\rm int},m^2}(T)$ 
of the magnetic autocorrelation function $A_{m^2}(t)$ over the time
interval $0<t<T$. These results apply to the single-cluster simulation
of the $q=1.25$ random-cluster model.}
\label{figmt1_25t}
\end{figure}

\begin{figure}
\begin{center}
\leavevmode
\epsfxsize 10.6cm
\epsfbox{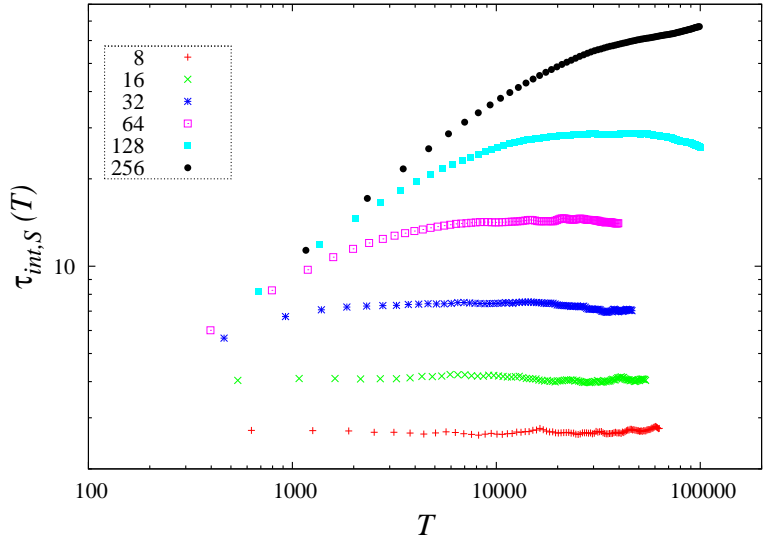}
\end{center}
\caption{(Color online) Indefinite integral $\tau_{{\rm int},S}(T)$ of
the autocorrelation function $A_{S}(t)$ for the single-cluster size over
the time interval $0<t<T$.  These results apply to the single-cluster
simulation of the $q=1.25$ random-cluster model.}
\label{figwf1_25t}
\end{figure}

The power-law dependence of the amplitude of the exponential decay of
autocorrelations means that, in terms of a measure of the efficiency
of the algorithm, the significance of the dynamic exponent
$z_{\rm exp}$ is limited. The exponent $z_{\rm int}$ is a better
measure of the  $L$-dependence of the rate of decay of correlations,
because it includes the
size-dependence of the amplitude of the decay. The unusual behavior
in Fig.~\ref{fig_gmt1_25a} may be expected to
lead to significant differences between $z_{\rm exp}$ and  $z_{\rm int}$.
This expectation is verified by integration of Eq.~(\ref{beh_rho_realq})
which yields that
\begin{equation}
\tau_{{\text{int}},\obs} \propto L^{z_{\rm int}} \, , \mbox{\hspace{8mm}}
z_{\rm int} = (1-r_{\obs})z_{\rm exp} -s_{\obs} \, .
\label{tauintanal}
\end{equation}
Inspection of the numerical results for $z_{\rm exp}$, $r_{\obs}$ and
$s_{\obs}$ in Table \ref{tab_realq} shows that $z_{\rm exp}$ and 
$z_{\rm int}$ must have different values.
For a numerical analysis of $z_{\rm int}$, we have,
in line with Eq.~(\ref{tauint_definition}), integrated the
autocorrelation functions for $m^2$ and $S$ according to 
\begin{equation}
\tau_{{\text{int}},\obs}(T) \equiv \frac{1}{2}+\sum_{t=1}^{T}\,
A_{\obs }(t) \, ,
\label{tauintnum}
\end{equation}
where, presently, $T$ assumes the meaning of a time variable.
The integrated autocorrelation times of the $q=1.25$ model for
$\tau_{{\rm int},m^2}(T)$ and $\tau_{{\rm int},S}(T)$ are shown in
Figs.~\ref{figmt1_25t} and \ref{figwf1_25t}  respectively. The lines for 
large $L$ are approximately straight, which reflects the algebraic
decay of the autocorrelation functions as a function of $t$. As a
consequence of the exponential decay at large $t$, $\tau_{{\rm int}}(T)$
approaches a constant. However, integration of random correlations at
large $t$ eventually affects the accuracy of the numerical result for
$\tau_{{\rm int}}(T)$ so that a cutoff has to be applied for optimal
results.
For this reason, the integrated autocorrelation times for $L=256$  
could not be accurately determined and were skipped
from the analysis. The remaining data were fitted by
\begin{equation}
\tau_{{\rm int},O}(L) = A +B L^{z_{{\rm int},O}} \; ,
\label{fit_tint}
\end{equation}
where $A$ and $B$ are unknown constants. The fits for $q=1.25$ yield
$z_{{\rm int},m^2} = 1.4~(1)$, and $z_{{\rm int},S} = 1.1~(1)$.
Table \ref{tab_zint} includes the results for the $z_{{\rm int},\obs}$
for several other values of $q$.


\section {Discussion }
\label{sec4}

As stated in Sec.~\ref{sec1}, one might expect that the present
single-cluster algorithm would have a dynamic exponent that is about the 
same as that of the SWCM algorithm~\cite{CM}. However, after comparing
the dynamic exponents of both algorithms, we find that this expectation
is not justified for noninteger $q$. The single-cluster algorithm
formulated in this work represents a new dynamic universality class.
Finding the reasons behind this curious fact should help us to better 
understand from where critical slowing down arises, and tell us 
something about how one can further develop efficient Monte Carlo
algorithms in statistical physics.

The single-cluster algorithm described above is obviously related to the
Wolff \cite{UW} algorithm as defined for integer-$q$ Potts models;
it can reduce to the Wolff method if $q$ is an integer.
On the other hand, it is different in the sense 
that the single-cluster algorithm acts on a mixed configuration
of site variables and random-cluster variables. 

This mixture of different types of variables is essentially the
reason that the present single-cluster algorithm is relatively slow.
In this algorithm, a number of lattice sites belongs to random 
clusters of type 0 with weight $q-[q]$, while the remaining sites
are decorated with a Potts variable in one of $[q]$ Potts states.

\begin{figure}
\begin{center}
\leavevmode
\epsfxsize 10.6cm
\epsfbox{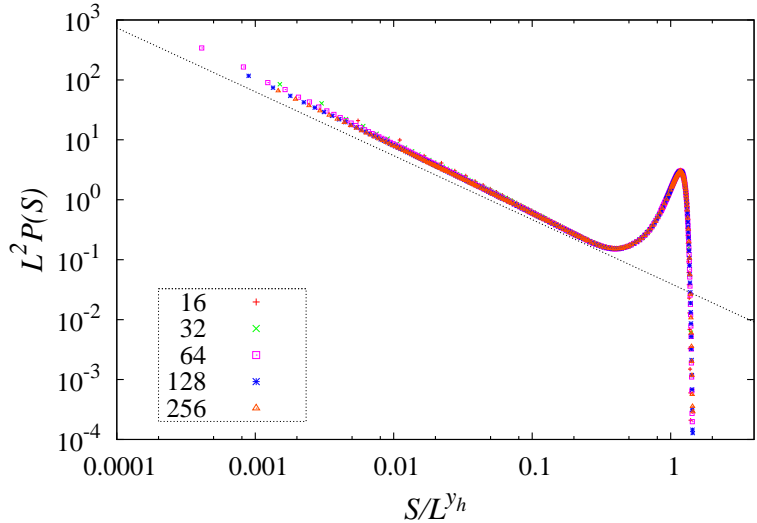}
\end{center}
\caption{(Color online) Data collapse for the single-cluster
distribution $P(S)$ as a
function of the cluster size $S$ for the critical $q=2$ random-cluster
model. The dashed line illustrates the asymptotic slope $-2/y_h=16/15$
which applies to $1<<S<<L^{y_h}$. Data are shown for system sizes
$L=16$, 32, 64, 128 and 256. The quantity $P$ represents the probability
that a randomly chosen site belongs to a cluster of size $S$.
}
\label{fig_q2_scld} 
\end{figure}

As described in Sec.~\ref{sec2}, the only process that can change a 
type-0 cluster back into an integer spin state, depends on the random
selection of a site in that cluster in the beginning of each cluster
step.  Thus, large clusters of type 0 are short-lived and small ones
are long-lived. It is illustrated in Fig.~\ref{fig_q2_scld} that the
single-cluster distribution for the case $q=2$ displays a wide range
of algebraic decay and an additional maximum at large cluster sizes of
order $L^{y_h}$, preceding a rapid decay at even larger sizes.
The distribution shown in Fig.~\ref{fig_q2_scld} represents a time
average. Individual cluster decompositions deviate because of thermal
fluctuations. The lifetime of these deviations will naturally depend
on the cluster size. The smaller the type-0 clusters are, the longer 
they will persist, and this will be reflected  in the decay of the 
autocorrelation functions. The pronounced maximum in Fig.~\ref{fig_q2_scld}
at $S\approx L^{y_h}$ can thus be associated with a rapid initial decay
of the autocorrelations. Once the largest clusters of type 0 are updated,
some autocorrelations are still persisting due to the thermal fluctuations
of the numbers of smaller clusters that remain to be updated.  After $t'$
single-cluster steps, the autocorrelations of the numbers of clusters with
sizes $S>L^2/t'$ will be strongly reduced, while the clusters with sizes
$S<L^2/t'$ will mostly be unaffected. Since the cluster-size distribution
decays algebraically in a range of $S$, it is natural that autocorrelations
associated with clusters that are not yet updated display a corresponding
power-law decay in time, as long as the smallest clusters survive.
After a number of steps of order $L^2$ also the clusters of size 1 will
be updated. This somewhat qualitative reasoning, which neglects any
persisting correlations after all clusters are visited, would mean that
the longest autocorrelation time, expressed in single-cluster updates,
is of order $L^2$, after which
the autocorrelations will decay exponentially. Expressed in units of $t$
as defined in Sec.~\ref{autocorrelation times}, this corresponds with
autocorrelations scaling as $L^{2y_h-2}$ at criticality. Our numerical
results suggest that the dynamic exponent is slightly larger, namely
$z_{\rm exp}\approx 2$, but the data do not allow a more firm statement.

The persistence of the smallest clusters during a time of approximate
order $L^2$ leads to a long time "tail" during simulations using the
single-cluster method. It is this effect that we hold responsible for the
relatively large dynamic exponent $z_{\rm exp}$ of the single-cluster
method. However, the amplitude of the algebraic decay of the 
autocorrelation functions still depends with a factor $ L^{-s_{\obs}}$
on the system size $L$.  A positive value of $s_{\obs}$ therefore means
that the critical slowing down is less severe than suggested by the
value of $z_{\rm exp}$, in agreement with the smaller values of
$z_{\rm int}$ as shown in Table \ref{tab_zint}.

Nevertheless our findings indicate that the single-cluster algorithm,
apart from displaying interesting dynamic behavior, is not an
efficient tool to investigate the two-dimensional random-cluster model.
In higher-dimensional systems we have similar expectations. But there
still seems to be a possibility that a number of single-cluster steps 
alternating with a full-cluster decomposition, which takes advantage
of the fast initial decay of autocorrelations of the single-cluster
algorithm as well as of the absence of a long-time tail in the
SWCM cluster algorithm, will be relatively efficient in
higher-dimensional systems.

\acknowledgments
We are indebted to J. R. Heringa and J.-S. Wang for valuable discussions.
We acknowledge the hospitality of the Institute for Mathematical Sciences
of the National University of Singapore (2004), where we learned about
the status of continuous-$q$ Monte Carlo algorithms.  This research
is partly supported by the Science Foundation of the Chinese
Academy of Sciences, and by the Program for New Century Excellent Talents
in Chinese Universities (NCET).

%

\vfill
\newpage

\begin{table}
\caption{Simulation results for the average squared magnetization
$\langle m^2\rangle$ and the single-cluster size ${S}$ for the critical
$q=2$ random-cluster model, as obtained by the Wolff (W) and the present
single-cluster algorithm (S) with $n=q-1$ as defined in the text.
The parameter $L$ specifies the linear system size. The number of
samples per system size is $4 \times 10^6$ for each simulation, and 
the number of clusters formed between subsequent samples is $2$ 
for $L \leq 24$ and $3$ for $L=32$. The numbers between brackets 
show the statistical error margins in the last two decimal places.}

\label{tab_1}  
\begin{center}
\begin{tabular}{|c|llllll|}
\hline
$L $         & ~~8             & ~~12            & ~~16 
             & ~~20            & ~~24            & ~~32          \\ 
\hline
$ m^2 $ (W)  & 0.64693 (18)    & 0.58581 (18)    & 0.54537 (17)   
             & 0.51584 (16)    & 0.49305 (17)    & 0.45874 (14)  \\ 
$ m^2 $ (S)  & 0.6478~ ~(6)    & 0.5861~ ~(8)    & 0.5442~ ~(9)   
             & 0.5164~ (10)    & 0.4932~ (13)    & 0.4610~ (12)  \\ 
\hline
${S}  $ (W)  & 0.64666 (18)    & 0.58581 (18)    & 0.54544 (17)   
             & 0.51594 (16)    & 0.49311 (17)    & 0.45878 (14)  \\ 
${S}  $ (S)  & 0.6470~ ~(6)    & 0.5860~ ~(8)    & 0.5441~ ~(9)   
             & 0.5165~ (10)    & 0.4932~ (13)    & 0.4610~ (12)  \\ 
\hline
\end{tabular}
\end{center}
\end{table}    

\begin{table}
\caption{Lengths of the Wolff-type simulations in Sec.~\ref{intq}
for $ L\geq 16$ and $q=2$, 3, and 4, in units of $10^7$ samples.}
\label{tab_sample}
\begin{center}
\begin{tabular}{|r|rrrrrrrrr|}
\hline
 $ L $      &~~~16    &~~~32      &~~~64      &~~128  
&~~256      &~~512    &~~1024     &~2048      &~4096   \\
\hline
$ q=2 $     & 12      & 12        & 12        & 12     
& 16        & 16      & 32        & 8         & 8       \\
$ q=3 $     & 4       & 4         & 4         & 4      
& 8         & 8       & 12        & 12        & --      \\
$ q=4 $     & 8       & 12        & 20        & 32     
& 48        & 72      & 64        & --        & --      \\
\hline
\end{tabular}
\end{center}
\end{table}

\begin{table}
\caption{Single-cluster dynamics for several values of $q$.
The exponents $r_s$, and $r_m$ are those in
Eq.~(\ref{beh_rho_realq}) for ${S}$, and $m^2$, respectively; 
the same labeling applies to $s_s$ and $s_m$. The values of $s_m$
are not significantly different from zero for noninteger values of $q$. 
For the purpose of comparison, the last column shows results~\cite{Deng}
for $z_{\rm exp}$ applying to SWCM cluster dynamics. Furthermore, some
data are included for integer values $q=2$, 3 and 4, with the choice
$n=q$, so that these results apply to the Wolff algorithm.}
\label{tab_realq}
\begin{center}
\begin{tabular}{|l|lllll|l|}
\hline
~~~$q$~~~~&~~$s_s$~~&~~$r_s$~~&~~$s_m$~~&~~$r_m$~~&~~$z_{\rm exp}$~~&
                                                     $z_{\rm exp}$ (SW)\\
\hline
~1.25&~0.25~(2) &~0.25~(2)&~0.00~(2) &~0.25~(1)&~2.0~(2)   &~0.00      \\
~1.50&~0.19~(2) &~0.19~(2)&~0.00~(2) &~0.19~(1)&~2.0~(2)   &~0.00      \\
~1.75&~0.14~(2) &~0.14~(2)&~0.00~(2) &~0.14~(1)&~2.0~(2)   &~0.06 ~(1) \\
~2.25&~0.26~(2) &~0.15~(2)&~0.00~(2) &~0.14~(1)&~2.0~(2)   &~0.24 ~(1) \\
~2.50&~0.22~(2) &~0.10~(2)&~0.00~(2) &~0.12~(1)&~2.0~(2)   &~0.31 ~(1) \\
~2.75&~0.17~(2) &~0.06~(2)&~0.00~(2) &~0.10~(1)&~2.0~(2)   &~0.42 ~(2) \\
\hline
~2.00&~0.37~(2) & ~~--    &~0.14~(2) & ~~--    &~0.07~ (1) &~0.14~(1)  \\
~3.00&~0.34~(2) & ~~--    &~0.05~(2) & ~~--    &~0.521 (7) &~0.49~(1)  \\
~4.00&~0.25~(2) & ~~--    &~0.00~(2) & ~~--    &~1.007 (9) &~0.93~(2)  \\
\hline
\end{tabular}
\end{center}
\end{table}

\begin{table}
\caption{Dynamic exponent $z_{{\rm int}}$ of the single-cluster
cluster algorithm. This exponent describes the scaling behavior of
$\tau_{\rm int}$, the integrated autocorrelation function.
For a  negative exponent $z_{{\rm int}}$, the $\tau_{\rm int}$
data approach a constant when $L \rightarrow \infty$.
The values of $z^*_{{\rm int},\obs}$ are calculated from
Eq.~(\ref{tauintanal}) and Table~\ref{tab_realq}, while those of
$z_{{\rm int},\obs}$ follow from the fits using Eq.~(\ref{fit_tint}).
Some data are included for integer $q$; these
results apply to the Wolff algorithm. }
\label{tab_zint}
\begin{center}
\begin{tabular}{|l|llllll|lll|}
\hline
 $q$   & $1.25$~~~~~  & $1.50$~~~~~  & $1.75$~~~~~
       & $2.25$~~~~~  & $2.50$~~~~~  & $2.75$~~~~~
       & ~~$2$~~~     & ~~$3$~~~     & ~~$4$    \\
\hline
 $z_{{\rm int},m^2}$
       & $1.4~(1)$     & $1.5 ~(1)$     & $1.6~ (1)$
       & $1.9~(1)$     & $1.9 ~(1)$     & $2.0~ (1)$
       & $-0.16~(2)$   & $0.485~(7)$    & $1.005~(9)$ \\
 $z^*_{{\rm int},m^2}$
       & $1.5~(2)$     & $1.6 ~(2)$     & $1.7~ (2)$
       & $1.7~(2)$     & $1.8 ~(2)$     & $2.0~ (2)$
       & --            & --             & --         \\
\hline
 $z_{{\rm int},S}$
       & $1.1~(1)$     & $1.1 ~(1)$     & $1.3~ (1)$
       & $1.6~(1)$     & $1.7 ~(1)$     & $1.8~ (1)$
       & $-0.4~(1)$    & $0.16~(4)$     & $0.72~(5)$      \\
 $z^*_{{\rm int},S}$
       & $1.3~(2)$     & $1.4 ~(2)$     & $1.6~ (2)$
       & $1.4~(2)$     & $1.6 ~(2)$     & $1.7~ (3)$
       & --            & --             & --         \\
\hline
\end{tabular}
\end{center}
\end{table}

\end{document}